\documentclass[runningheads]{llncs}
\usepackage{graphicx}
\usepackage{bm}
\usepackage{subcaption}
\usepackage{svg}
\usepackage{soul}
\usepackage{hyperref}
\usepackage{amsmath}
\usepackage{amsfonts}
\usepackage{amssymb}

\begin{document}
\title{The Challenge of Fetal Cardiac MRI Reconstruction Using Deep Learning}
%
%

%
%
\author{Denis Prokopenko\inst{1} \and
Kerstin Hammernik\inst{2,3} \and
Thomas Roberts\inst{1,4} \and
David F A Lloyd\inst{5,6} \and
Daniel Rueckert\inst{2,3} \and
Joseph V Hajnal\inst{1}}

\authorrunning{D. Prokopenko et al.}

\institute{Biomedical Engineering Department, School of Biomedical Engineering \& Imaging Sciences, King's College London, London, UK\and
Department of Informatics, Technical University of Munich, Munich, Germany \and
Department of Computing, Imperial College London, London, UK \and
Clinical Scientific Computing, Guy's \& St. Thomas' NHS Foundation Trust, London, UK \and
Child Health, King's College London, London, United Kingdom  \and
Paediatric and Fetal Cardiology, Evelina London Children's Hospital, London, UK\\
\email{denis.prokopenko@kcl.ac.uk}}
\maketitle              
\setcounter{footnote}{0}
%

%
%
\begin{abstract}

Dynamic free-breathing fetal cardiac MRI is one of the most challenging modalities, which requires high temporal and spatial resolution to depict rapid changes in a small fetal heart.
The ability of deep learning methods to recover undersampled data could help to optimise the kt-SENSE acquisition strategy and improve non-gated kt-SENSE reconstruction quality.
In this work, we explore supervised deep learning networks for reconstruction of kt-SENSE style acquired data using an extensive in vivo dataset.
Having access to fully-sampled low-resolution multi-coil fetal cardiac MRI, we study the performance of the networks to recover fully-sampled data from undersampled data.
We consider model architectures together with training strategies taking into account their application in the real clinical setup used to collect the dataset to enable networks to recover prospectively undersampled data.
We explore a set of modifications to form a baseline performance evaluation for dynamic fetal cardiac MRI on real data.
We systematically evaluate the models on coil-combined data to reveal the effect of the suggested changes to the architecture in the context of fetal heart properties.
We show that the best-performers recover a detailed depiction of the maternal anatomy on a large scale, but the dynamic properties of the fetal heart are under-represented.
Training directly on multi-coil data improves the performance of the models, allows their prospective application to undersampled data and makes them outperform CTFNet introduced for adult cardiac cine MRI.
However, these models deliver similar qualitative performances recovering the maternal body very well but underestimating the dynamic properties of fetal heart.
This dynamic feature of fast change of fetal heart that is highly localised suggests both more targeted training and evaluation methods might be needed for fetal heart application.

\end{abstract}

\section{Introduction}\label{sec:intro}

Dynamic free-breathing fetal cardiac imaging is one of the most challenging applications of magnetic resonance imaging (MRI).
Fetal heart imaging requires a large field of view to encompass the maternal anatomy combined with high spatial and temporal resolutions to depict tiny dynamic structures of the fetal heart.
The imaging must accommodate uncontrolled fetal motion and maternal respiration while capturing fetal cardiac beating, which could be more than twice as fast as an adult heartbeat.

\begin{figure}[t]
    \begin{minipage}[b]{0.5\textwidth}
    \end{minipage}
    \hfill
    \begin{minipage}[b]{0.35\textwidth}
        \centering
        \includegraphics[width=\textwidth]{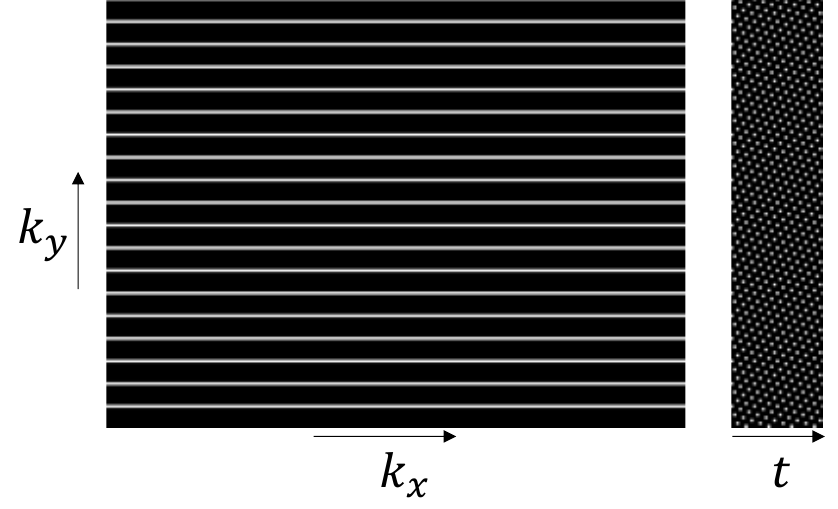}
        \subcaption{}\label{fig:undersampled}
    \end{minipage}
    \hfill
    \begin{minipage}[b]{0.35\textwidth}
        \centering
        \includegraphics[width=\textwidth]{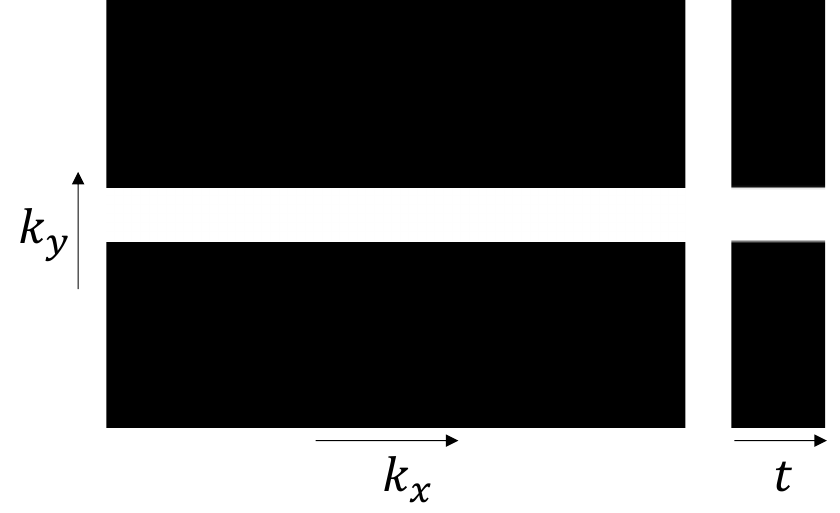}
        \subcaption{}\label{fig:dense}
    \end{minipage}
    \hfill
    \begin{minipage}[b]{0.5\textwidth}
    \end{minipage}
    \caption{The kt-SENSE acquisition patterns to sample high-resolution undersampled (a) and low-resolution densely sampled (b) \textit{kt} data.}
    \label{fig:sampling}
\end{figure}

Conventional kt-SENSE~\cite{tsao2003k} can deliver dynamic MRI of fetal heart across multiple cardiac cycles using a prior-led reconstruction of highly undersampled signals.
While there are other reconstruction methods to deliver dynamic fetal cardiac MRI using non-Cartesian or non-uniform sampling~\cite{jung2009k,lingala2011accelerated} or gating~\cite{haris2017self,kording2018dynamic}, we focus on the kt-SENSE approach as it was used to acquire the substantial dataset of fetal cardiac MRI.
The kt-SENSE is simpler to use to collect the data compared to non-uniform or non-Cartesian methods and provides real-time fetal cardiac data over multiple cardiac cycles without averaging in contrast to gated approaches. 
In case of kt-SENSE, the data is usually acquired in two stages using sampling patterns, which have high temporal resolution to freeze bulk fetal and fetal cardiac motion.
One acquisition pattern in Figure \ref{fig:undersampled} samples high-resolution undersampled \textit{k}-space data in time (\textit{kt} data). 
The other pattern in Figure \ref{fig:dense} acquires low-resolution (low \textit{k}) fully-sampled \textit{kt} data, which helps to disambiguate dynamic features in undersampled data during reconstruction.
However, the two-stage acquisition approach is inefficient because the same anatomy is sampled twice, while changes in fetal pose between the two samplings could introduce differences in the collected data reducing the quality of the reconstruction.

Recent advances in deep learning (DL) for dynamic adult cardiac MRI reconstruction~\cite{desai2021noise2recon,qin2021complementary,qin2018convolutional,qin2019k,schlemper2017deep} expanded the variety of methods that can be used for the reconstruction.
The ability to recover fully-sampled data directly from undersampled information is a common aim of the proposed DL reconstruction methods.
In the case of fetal cardiac MRI, they have the potential to replace the dense acquisition part (Figure \ref{fig:dense}) with a reconstruction from data acquired with undersampling pattern (Figure \ref{fig:undersampled}).
As a result, it could reduce the acquisition time up to two times, using only one part of the two-stage sampling approach, and increase the robustness of the reconstruction to the spontaneous motion between the stages of acquisition.
However, the use of DL reconstruction methods for fetal cardiac MRI reconstruction is limited due to additional constraints of the application domain.
The fully-sampled high-resolution fetal cardiac ground truth data is currently unavailable.
This is due to the limitations on the spatio-temporal resolution required to freeze the motion and scanner acquisition rate.
Also, the lack of real high-resolution dynamic MRI data forces DL methods to employ high-resolution reconstructions as ground truth.
As a result, the models are optimised to mimic the target reconstruction algorithms with their biases rather than the true depiction of the anatomy.
In addition, the networks usually perform dynamic adult heart MRI reconstruction from retrospectively undersampled gated cine modality, which differs from the fetal cardiac case, where the recovery of the heartbeat over multiple cycles is desired.
In the absence of the ground truth data, emerged self-supervised versions showed similar performance to supervised models according to the numerical evaluations~\cite{acar2021self,zou2022selfcolearn}.
However, the self-supervised models delivered a less reliable depiction of the moving edges, which are the most sensitive anatomy in the case of fetal cardiac MRI.

In this work, we study the performance of supervised DL baselines to recover fully-sampled low-resolution dynamic fetal cardiac MRI from undersampled data to make the kt-SENSE~\cite{tsao2003k} algorithm more robust to motion and shorten the acquisition time.
First, we present and compare a set of U-Net-based architectures trained on data combined over multiple coils (coil-combined) to recover undersampled data, focusing on dynamic properties of fetal heart.
Next, we take the best-performing versions and present their performance being trained on multi-coil data, which enables their prospective application to the existing kt-SENSE acquisition and reconstruction pipelines and direct comparison with the state-of-the-art Complementary Time-Frequency domain Network (CTFNet) proposed for gated adult heart MRI reconstruction.

\section{Methods}\label{sec:methods}

We investigate the reconstruction of free-breathing dynamic fetal heart MRI volumes acquired with a multi-coil receiver array as a part of \textit{kt}-SENSE~\cite{tsao2003k} algorithm.
We use the fully-sampled low-resolution data acquired from scanner to study DL methods to recover real depiction of anatomy rather than mimicking conventional reconstruction algorithms with their biases.
The data is sampled in \textit{kt} space, but it can be represented in temporal image space (\textit{xt}) and temporal frequency image space (\textit{xf}) as well.
First, we use the coil-combined SENSE~\cite{pruessmann1999sense} data to study the reconstruction quality of dynamic features of fetal heart using a set of modified U-Net models~\cite{ronneberger2015u}.
Next, we take the best-performing models and train them directly on multi-coil data to use the advantages of the diverse coil data.
Training on coil data allows us to optimise the models in a clinically relevant way to work with the undersampled data acquired prospectively from the scanner.
Finally, we compare the models with an iterative Complementary Time-Frequency Domain Network (CTFNet)~\cite{qin2021complementary}, which was introduced as a state-of-the-art method for adult cine MRI reconstruction.

\begin{figure}[t]
    \centering
    \begin{minipage}[b]{0.8\textwidth}
        \centering
        \includegraphics[width=\textwidth]{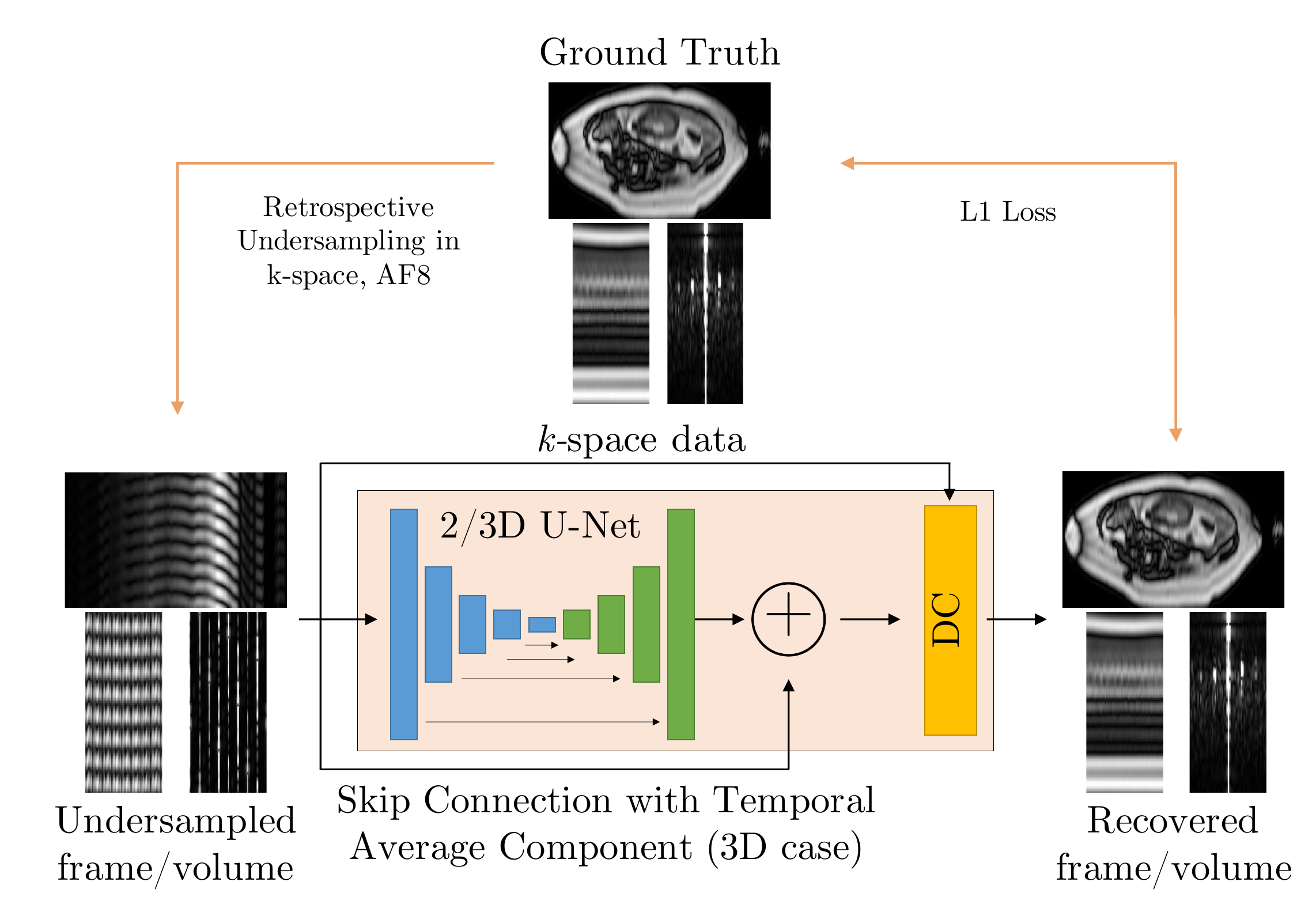}
    \end{minipage}
    \caption{The architecture with optional data consistency (DC) layer~\cite{schlemper2017deep} and skip connection propagating temporal average. The skip connection is used for 3D models trained on data in the \textit{xf} domain.}
    \label{fig:model}
\end{figure}

The 2D version of the U-Net based model with optional data consistency (DC) layer~\cite{schlemper2017deep} in Figure \ref{fig:model} processes each complex-valued image frame in \textit{xt} space independently without explicitly exploiting the spatio-temporal correlations available in fetal cardiac data.
To enrich the input with temporal information, we add a sliding window average or a temporal average to the undersampled data as additional input channels.
The sliding window average is a sum of $R$ consecutive image frames, where $R=8$ is the acceleration factor of undersampling.
Temporal averaging delivers an average of all frames scaled by acceleration factor $R$.

The 3D version of the U-Net based model with an optional DC layer~\cite{schlemper2017deep} in Figure \ref{fig:model} processes the volumes that contain all available temporal information within the slice volume.
In contrast to the 2D case, the 3D models can process data in the temporal frequency (\textit{xf}) domain, which is a sparser representation than the \textit{xt} representation and usually helps to improve the reconstruction quality \cite{lustig2007sparse,tsao2003k}.
Since the \textit{xf} representation has most of the energy concentrated in the temporal average frame, the optimisation of the loss function is likely to favour reconstruction of the temporal average rather than the lower energy dynamic temporal frequencies. 
To shift the focus to non-zero frequency frames, we add the temporal average to the 3D U-Net output via a skip connection shown in Figure \ref{fig:model}.

Our fetal cardiac MRI dataset includes multi-planar stacks of multi-slice multi-coil dynamic fetal cardiac MRI scans acquired from 56 subjects (3 healthy volunteers, 53 patients with congenital heart diseases, $55809$ dynamic 3D coil volumes).
The bSSFP sequence~\cite{carr1958steady} was used together with regular Cartesian \textit{kt} undersampling~\cite{tsao2005optimizing}, voxel size $2.0 \times 2.0 \times 6.0$ mm, $8\times$ acceleration, and 28 coil receiver to acquire the data on a Philips Ingenia 1.5T MR system.
Ethical approvals were obtained (REC 14/LO/1806, REC 07/H0707/105), and each subject gave informed consent.
A subset of $20\%$ of cases chosen on a patient level is used for testing only.
The rest of the data is used for training the models with $20\%$ of it set apart for validation.
The low-resolution fully-sampled signal is sampled across $19$ central \textit{k}-space lines and zero-padded in the phase-encoding direction producing images of resolution $152 \times 400$ for $64$ frames, with a temporal resolution of $72$ ms per frame, which is sufficient to capture fetal cardiac motion~\cite{roberts2020fetal}.
In addition, coil sensitivity maps and coil noise covariance matrix were acquired for each multi-coil volume, which we use to normalise the data to the unit noise level.
While the coil data forms our multi-coil dataset, we create an additional coil-combined dataset by combining coil data using SENSE reconstruction~\cite{pruessmann1999sense}.
To mimic the kt-SENSE~\cite{tsao2003k} acquisition sampling used to collect the dataset, we retrospectively undersample data with the uniform lattice pattern~\cite{tsao2005optimizing} and acceleration factor $R = 8$.

The U-Net architectures follow the original implementation~\cite{ronneberger2015u} with convolutional kernel size $3$ and $4$ downsampling steps with $2\times 2$ max pooling.
The optional data consistency (DC) layer has three modes: no DC, forced DC and adjustable DC.
The forced option replaces the output of the model with available input data, while the adjustable option uses a merging coefficient which is optimised as an additional model parameter.
For both datasets, the U-Net modifications were trained to optimise the $\mathcal{L}_1$ loss function for $50$ epochs using Adam~\cite{kingma2014adam} with a learning rate of $0.0001$ and exponential decay of $0.95$ using the PyTorch framework~\cite{paszke2019pytorch}.
The final set of model parameters is chosen based on the performance of the models on the validation subset.

In the case of CTFNet, the architecture follows the original implementation and training procedure~\cite{qin2021complementary}. The model was tuned over $150$ epochs optimising the $\mathcal{L}_1$ loss function on the patches using Adam with the same learning rate and decay as for U-Net-based architectures.
In addition, we used adjustable DC coefficient to merge the available input information into model prediction.

The evaluation is performed for $14$ modifications of U-Net trained on the coil-combined dataset and for the $3$ best-performing versions of U-Net together with CTFNet trained on multi-coil data.
The quantitative evaluation compares coil-combined predicted volumes and ground truth data in the \textit{xt} domain per volume according to normalised mean squared error (NMSE), mean squared error (MSE), mean absolute error (MAE), peak signal-to-noise ratio (PSNR) and structural similarity (SSIM) over a masked body region. 
While the NMSE, MSE and MAE compare tensors with complex intensities, magnitude tensors are evaluated using PyTorch Image Quality (PIQ)~\cite{kastryulin2022pytorch} implementation of PSNR and SSIM~\cite{piq}.
The presence of important dynamic features of fetal heart is evaluated visually in temporal and frequency domains.

\section{Results}\label{sec:results}
\begin{table}[t]
    \centering
    \caption{Comparison of the U-Net models trained on coil-combined dataset. 
    }
    \label{tab:label_sense}
    \begin{tabular}{|l |c c c c c|}
        \hline
        Model description & NMSE $\downarrow$& MSE $\downarrow$ & MAE $\downarrow$ & PSNR $\uparrow$ & SSIM $\uparrow$\\
        \hline
        2D U-Net, no DC & $0.7976$ & $4102.3838$ & $49.3082$ & $12.2359$ & $0.6209$ \\
        2D U-Net, adjustable (adj.) DC & $0.7829$ & $4019.8306$ & $49.0102$ & $12.3327$ & $0.6324$ \\
        2D U-Net, forced DC & $0.7715$ & $3967.0254$ & $49.1194$ & $12.3991$ & $0.6393$ \\
        \hline
        2D U-Net, sliding window & $0.0062$ & $32.0378$ & $4.1111$ & $33.7646$ & $0.9826$ \\
        2D U-Net, temporal average & $0.0176$ & $91.4503$ & $5.7255$ & $29.6221$ & $0.9658$ \\
        \hline
        3D U-Net, \textit{xt}, no DC & $0.0130$ & $67.1712$ & $5.7132$ & $30.2262$ & $0.9678$ \\
        3D U-Net, \textit{xt}, adj. DC & $0.0108$ & $55.7603$ & $5.3444$ & $31.0387$ & $0.9710$ \\
        3D U-Net, \textit{xt}, forced DC & $0.0107$ & $55.1941$ & $5.3293$ & $31.0834$ & $0.9713$ \\
        \hline
        3D U-Net, \textit{xf}, no DC & $0.0097$ & $50.7774$ & $4.8127$ & $31.6965$ & $0.9742$ \\
        3D U-Net, \textit{xf}, adj. DC & $0.0091$ & $47.1734$ & $4.7795$ & $32.0303$ & $0.9772$ \\
        3D U-Net, \textit{xf}, forced DC & $0.0087$ & $45.5490$ & $4.7054$ & $32.2168$ & $0.9774$ \\
        \hline
        3D U-Net, \textit{xf}, skip connect, no DC & $0.0053$ & $27.7157$ & $3.6231$ & $34.4853$ & $0.9854$ \\
        3D U-Net, \textit{xf}, skip connect, adj. DC & \bm{$0.0049$} & $25.5143$ & $3.5402$ & $34.8319$ & $0.9863$ \\
        3D U-Net, \textit{xf}, skip connect, forced DC & \bm{$0.0049$} & \bm{$25.3647$} & \bm{$3.5373$} & \bm{$34.8663$} & \bm{$0.9864$} \\
        \hline
    \end{tabular}
\end{table}

\begin{figure}[t]
    \centering
    \includegraphics[width=1\textwidth]{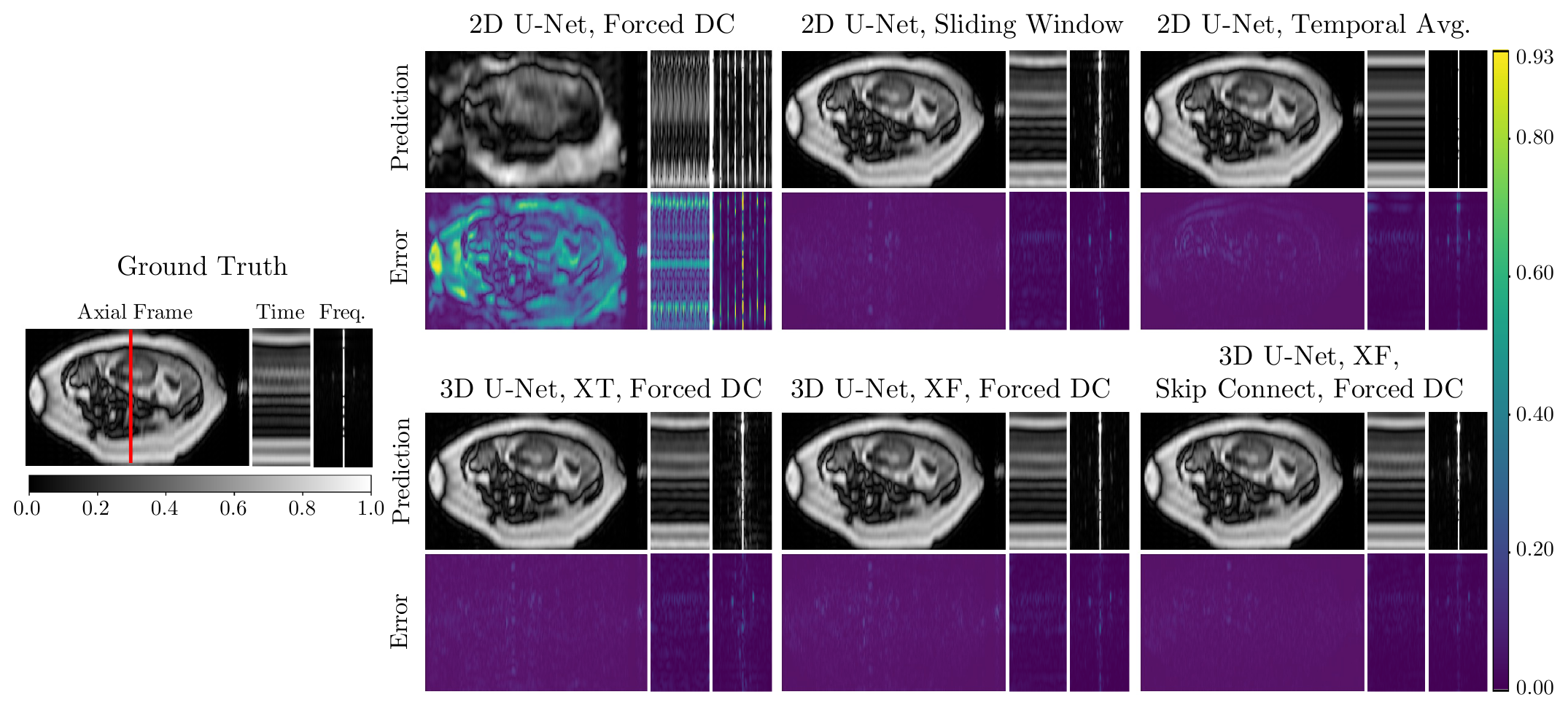}

    \caption{Comparison of the models trained on coil-combined dataset. The figure shows axial frame view and a slice through fetal heart (red line) in temporal (\textit{xt}) and frequency (\textit{xf}) domains for ground truth volume and each model prediction with the corresponding error maps.}\label{fig:train_sense}
\end{figure}

In case of model optimisation using coil-combined data, the best results were shown by 3D U-Nets with temporal average skip connection and data consistency trained using the \textit{xf} representation of data.
The model managed to deliver the best values according to evaluated metrics for both forced and adjustable modes of data consistency layer (Table \ref{tab:label_sense}).
The next best performance was achieved using the same model without data consistency.
Qualitatively, the models were highly effective at recovering accurate image depictions of static or slowly moving maternal and fetal anatomy as revealed by the subtraction images in Figure \ref{fig:train_sense}.
Even though the 3D U-Net with data consistency and skip connection delivered the closest to ground truth results for rapid changes in small regions such as fetal heart, there was still a clear under-representation of the beating of a tiny heart seen in both temporal and frequency domains.

Additional access to temporal information played a crucial role in improving the performance of the trained models.
In the simplest case of a 2D U-Net with DC, the frame-wise training approach fails to deliver reliable results with model overfitting on training data.
Temporal average and sliding window average information supplied with the input boosted the performance for all evaluated metrics compared to 2D U-Net without access to an additional input of temporal information.
The additional inputs helped to improve the 2D U-Net performance reducing the NMSE more than $100$ and almost $50$ times for sliding window and temporal average injections respectively and improving the SSIM to $0.9826$ and $0.9658$.
The same trend can be seen in Figure \ref{fig:train_sense} as the models are able to recover the main features of a slice, while 2D U-Net with sliding window resolved some slow dynamic features as well.

All variants of 3D U-Net showed improved performance compared to the 2D U-Net with DC according to evaluated metrics.
Qualitatively, 3D models recovered more information about anatomy movements and resolved aliasing artefacts slightly better.
The 3D U-Net case showed positive changes in model performance moving from a temporal representation of the data to a sparser \textit{xf} domain and additional propagation of temporal average via skip connection.
The combination of sparse data representation and skip connection helped to outperform 2D U-Net with sliding window, recovering a more detailed depiction of the fetal heart beating in both temporal and frequency domains (Figure \ref{fig:train_sense}).

\begin{figure}[t]
    \centering
    \includegraphics[width=\textwidth]{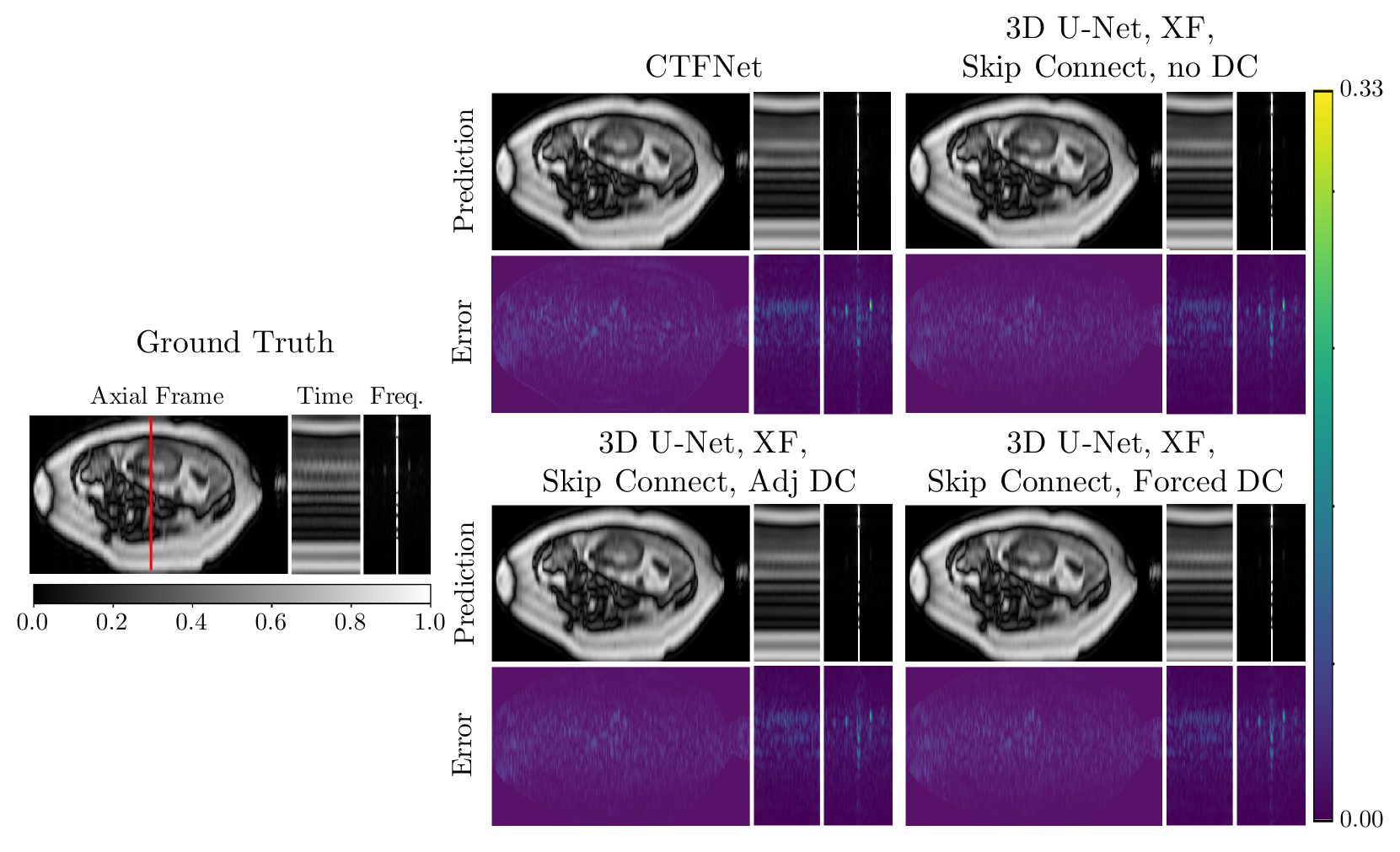}

    \caption{Comparison of the models trained on multi-coil dataset. The figure shows axial frame view and a slice through fetal heart (red line) in temporal (\textit{xt}) and frequency (\textit{xf}) domains for ground truth volume and each model prediction with the corresponding error maps.}\label{fig:train_coil}
\end{figure}
\begin{table}[t]
    \centering
    \caption{Comparison of the models trained on multi-coil dataset. 
    } \label{tab:coil_sense}
    \begin{tabular}{| l| c c c c c|}
        \hline
        Model description & NMSE $\downarrow$& MSE $\downarrow$ & MAE $\downarrow$ & PSNR $\uparrow$ & SSIM $\uparrow$ \\
        \hline
        CTFNet, adj. DC & $0.0082$ & $52.2186$ & $3.5009$ & $34.6011$ & $0.9869$ \\
        \hline
        3D U-Net, \textit{xf}, skip connect, no DC & \bm{$0.0032$} & $16.5106$ & $2.7226$ & $36.9108$ & $0.9897$ \\
        3D U-Net, \textit{xf}, skip connect, adj. DC & \bm{$0.0032$} & $16.3916$ & $2.7202$ & $36.9298$ & $0.9897$ \\
        3D U-Net, \textit{xf}, skip connect, forced DC &  \bm{$0.0032$} & \bm{$16.3363$} & \bm{$2.7088$} & \bm{$36.9645$} & \bm{$0.9898$} \\
        \hline
    \end{tabular}
\end{table}


The introduction of a data consistency layer enhanced the performance of all U-Net-based models and improved the evaluated metrics (Table \ref{tab:label_sense}).
The forced mode for DC enabled slightly better performance across the models than an adjustable injection of the available input into output.

In the case of multi-coil data, three versions of 3D U-Net in \textit{xf} domain with skip connection and CTFNet were fine-tuned and compared.
The 3D U-Net with skip connection in \textit{xf} space and forced DC delivered the best performance (Table \ref{tab:coil_sense}).
The 3D U-Net with adjustable and no DC showed the second and the third-best results, outperforming the state-of-the-art CTFNet model proposed for a superficially similar task of adult heart reconstruction.
The image representation of prediction delivered by the best-performing 3D U-Net has similar to the ground truth depiction of the anatomy in Figure \ref{fig:train_coil}.
However, the dynamic features of the recovered fetal heart remain underrepresented in both temporal and frequency domains.

\section{Discussion}\label{sec:discussion}

In this work, we studied the reconstruction quality of kt-SENSE acquired fetal heart MRIs introducing gradual improvements to the U-Net backbone model.
Most of the modifications delivered faithful reconstructions of the maternal anatomy and some dynamic features of adult breathing and fetal motion.
The best performance was delivered by the 3D U-Net with forced DC and skip connection optimised on \textit{xf} data representations.
The quantitative evaluation reported excellent performance according to estimated global image measures.
The reconstruction results showed a clear depiction of the maternal and main fetal features.
However, there is a room for further improvements to recover the highly-dynamic features of fetal heart.

The CTFNet performed less well than the best U-Net architectures by all measures.
It was able to recover the large-scale anatomy, but the periodic dynamics and tiny fetal heart size limited the reconstruction of fetal heartbeat.

Notable performance was shown by the 2D U-Net with sliding window, which outperformed the basic 3D U-Net, even though the latter should have benefited from access to all available frames.
It is evident that the sliding window average is highly effective at resolving the aliasing artefacts preserving slower time variations.
Still, the reconstructed details could not fully represent the fetal heartbeat.

In this work, we considered training the models to recover real-time fully-sampled low-resolution ground truth from data undersampled with uniform Cartesian undersampling pattern and acceleration factor $8$.
Such data strategy allows us to incorporate the presented models trained on coil data directly into existing fetal cardiac MRI pipelines based on kt-SENSE acquisition used to sample the dataset.
In addition, the use of fully-sampled low-resolution data helps to avoid biases that models could grasp from learning high-resolution reconstruction delivered by conventional methods.
While one could use other acquisition planning with non-uniform non-Cartesian patterns, lower acceleration factor or data binning to deliver better reconstruction quality~\cite{hammernik2018learning,kofler2019spatio,qin2021complementary,qin2019k}, the produced results would be out of touch with the acquisition procedure used to collect the dataset used in our study.

The presented results showed that the metrics helped to evaluate the global reconstruction quality.
However, they were less effective in describing the errors that only occur in small spatial regions.
Therefore, we believe that more localised metrics together with more advanced backbone architectures could push further the reconstruction quality of fetal heart delivered by the DL models.
While the local measures could help to address the unbalanced dynamic features in data, a more complex backbone could include explicit features focused on motion reconstruction.

\section{Conclusion}\label{sec:conclusion}

In this work, we studied the challenge of fetal cardiac MRI reconstruction using deep learning models.
We progressively improved the performance of the U-Net backbone by introducing a set of modifications, which resulted in a systematic comparison of $14$ versions of the model and the cutting-edge CTFNet, designed for adult cardiac imaging.
The most effective model was 3D U-Net with data consistency and temporal average skip connection trained in \textit{xf} domain according to both qualitative and quantitative assessments.
The CTFNet model designed for adult imaging data was not that effective for fetal cardiac reconstruction, highlighting the more challenging nature of the fetal heart compared to the adult heart.
The best-performing networks were highly effective at recovering the large-scale field of view.
However, they were unable to fully recover the detailed dynamic features of the fetal heart.
This feature of fast change that is highly localised suggests both more targeted training and evaluation methods might be needed for fetal heart application.
Exploring these will be the subject of future work.

\section{Acknowledgements}

We appreciate the funding from EPSRC Centre for Doctoral Training in Smart Medical Imaging EP/S022104/1, support from Philips Medical Systems and core funding from the Wellcome/EPSRC Centre for Medical Engineering WT 203148/Z/16/Z and NIHR Biomedical Research Centre at Guy’s and St Thomas’ NHS Trust.

%
%

\bibliographystyle{splncs04}
\bibliography{references}

\end{document}